\documentclass[a4paper,fleqn]{cas-dc}

\usepackage[authoryear]{natbib}
\usepackage{rotating}
\usepackage{pdflscape}
\usepackage{geometry}
\usepackage{algorithm} 
\usepackage{algpseudocode} 
\usepackage{tabularx}
\usepackage{booktabs}
\usepackage{soul}
\def\tsc#1{\csdef{#1}{\textsc{\lowercase{#1}}\xspace}}
\tsc{WGM}
\tsc{QE}

\newdefinition{rmk}{Remark}
\newproof{pf}{Proof}
\newproof{pot}{Proof of Theorem \ref{thm}}

\begin{document}
\let\WriteBookmarks\relax
\def\floatpagepagefraction{1}
\def\textpagefraction{.001}

\shorttitle{Exploring Community's Perception in Public Safety}    

\shortauthors{Rahimi}  

\title [mode = title]{Exploring Public's Perception of Safety and Video Surveillance Technology: A Survey Approach}  



\author{Babak Rahimi~Ardabili}[orcid=0000-0002-3936-4552]

\cormark[1]


\ead{brahimia@charlotte.edu}



\affiliation{organization={University of North Carolina at Charlotte},
            addressline={9201 University City Blvd}, 
            city={Charlotte},
            citysep={}, 
            postcode={28223}, 
            state={NC},
            country={USA}}


\author{Armin Danesh~Pazho}
\author{Ghazal Alinezhad~Noghre}
\author{Vinit Katariya} 
\author{Gordon Hull} 
\author{Shannon Reid}
\author{Hamed Tabkhi}






\cortext[1]{Corresponding author}


\begin{abstract}
Addressing public safety effectively requires incorporating diverse stakeholder perspectives, particularly those of the community, which are often underrepresented compared to other stakeholders. This study presents a comprehensive analysis of the community's general public safety concerns, their view of existing surveillance technologies, and their perception of AI-driven solutions for enhancing safety in urban environments, focusing on Charlotte, NC. Through a survey approach, including in-person surveys conducted in August and September 2023 with 410 participants, this research investigates demographic factors such as age, gender, ethnicity, and educational level to gain insights into public perception and concerns toward public safety and possible solutions. Based on the type of dependent variables, we utilized different statistical and significance analyses, such as logit regression and ordinal logistic regression, to explore the effects of demographic factors on the various dependent variables. Our results reveal demographic differences in public safety concerns. Younger females tend to feel less secure yet trust existing video surveillance systems, whereas older, educated individuals are more concerned about violent crimes in malls. Additionally, attitudes towards AI-driven surveillance differ: older Black individuals demonstrate support for it despite having concerns about data privacy, while educated females show a tendency towards skepticism.
\end{abstract}



\maketitle

\section{Introduction}\label{int}

Public safety is a diverse issue that requires a comprehensive understanding of various stakeholders' perspectives and values. The study of stakeholders' perceptions plays a crucial role in developing effective strategies for enhancing community safety and well-being \citep{jones2023stakeholder}. Stakeholders, from local authorities to community members, possess unique insights and values that significantly influence public safety policies and practices. Understanding these diverse viewpoints is essential for creating inclusive and effective safety measures \citep{badiora2020stakeholders}.

The community is a pivotal player in shaping public safety dynamics among stakeholders. As the primary beneficiary of safety policies, the community's involvement in decision-making ensures that the measures adopted are effective and resonate with the community's needs and expectations \citep{participation2019}. The community's active participation contributes to a sense of ownership and responsibility towards maintaining public safety, fostering a collaborative environment between authorities and residents \citep{oludoye2021socio}. 

With the rise of technology, various technological solutions have been utilized to improve public safety and address safety issues. Technologies such as police body-worn cameras and passive video surveillance have been instrumental. Recently, the advent of Artificial Intelligence (AI) and its capability to process extensive datasets, identify intricate patterns, and facilitate swift, informed decisions shows potential to improve public safety \citep{allam2016impact, anttiroiko2014smart, ahmad2022data, pazho2023ancilia, ardabili2023understanding}. 

However, public safety issues and the application of technological solutions, such as AI, raise significant ethical and privacy concerns \citep{taylor2016surveillance, zhang2019ethical, ardabili2022AAAI}. Deploying these technologies in public spaces can lead to dilemmas surrounding individual rights and societal safety \citep{oruma2023security}. Decision-making and policy formulation in this domain necessitate a comprehensive understanding of various stakeholders' perspectives, including community members, law enforcement agencies, and legal experts \citep{roberts2020stakeholder, jones2021community}. This approach ensures that policies not only leverage technological advancements for public safety but also respect and protect individual needs and values \citep{williams2020ethical}.

Despite the acknowledged importance of the community's role, there has been a noticeable gap in adequately addressing this stakeholder group in previous studies. Many research efforts have predominantly focused on the perspectives of formal institutions and authorities, often overlooking the nuanced and critical insights that community members offer \citep{badiora2020stakeholders}. This oversight can lead to a disconnect between policy implementation and the actual needs and values of the community, potentially undermining the effectiveness of public safety initiatives and measures \citep{jones2023stakeholder}.

This paper emphasizes exploring and understanding community perceptions and values in public safety, shifting from top-down technological solutions to a more engaged, community-centric approach. This method, rooted in ethical considerations and participatory research principles, represents a departure from traditional crime-fighting strategies by prioritizing collaboration and co-creation with the community. Our goal is to foster a holistic and inclusive public safety strategy that integrates the voices of all stakeholders, particularly the community, into policy development and implementation \citep{oludoye2021socio}.

The substantial population growth within the Charlotte Metro Area, which has increased significantly from 200,000 residents in 1960 to approximately 2,267,000 in 2023\footnote{https://www.macrotrends.net/}, has posed a significant challenge in upholding public safety and addressing criminal activities \citep{scott2020problem}. This surging population, in conjunction with a concerning rise in the incidence of violent crimes from 4,891 cases per 100,000 people in 2010 to 7,400 per 100,000 in 2021\footnote{https://www.pewresearch.org/}, highlights the urgent requirement for fresh approaches to detect and prevent criminal acts effectively. In this context, it is reasonable to explore how people in Charlotte, notably one of the leading cities in the United States, perceive public safety and the potential solutions to address them. 

We formulated the following research questions based on our community-centric approach, where we engaged directly with the Charlotte residents.

\textbf{Research Questions:}
\begin{enumerate}
\item[RQ1:] Do public safety concerns differ across different demographic groups?
\item[RQ2:] Do people with different demographic characteristics perceive the effectiveness of existing video surveillance systems differently?
\item[RQ3:] Does demographic background affect people's perception of AI-driven video surveillance technology?
\end{enumerate}

The contributions of this research encompass both theoretical and practical dimensions:
\begin{itemize}
    \item Identifying and prioritizing specific safety concerns within the community, offering insights into areas requiring immediate attention.
    \item Examination of community perceptions regarding existing safety technologies, highlighting opportunities and risks for AI-driven solutions to address safety gaps.
    \item  Facilitation of direct community involvement in technology development, incorporating their concerns to create solutions aligned with community needs and values.
\end{itemize}

Our findings show that public safety concerns vary significantly across demographics: younger females perceive a lack of law enforcement and increased crime rates yet feel safer with current video surveillance systems. In contrast, older, more educated individuals express less concern about human trafficking and street incidents, focusing more on violent crimes in enclosed spaces like malls. Additionally, attitudes towards AI-driven video surveillance differ markedly, with older Black individuals valuing its use despite data concerns. At the same time, educated females show skepticism, highlighting a need for tailored approaches in public safety measures.

In the subsequent sections, we delve into the methodologies employed, present survey findings, analyze policy considerations, and articulate recommendations. Through this multifaceted exploration, we hope to contribute meaningfully to the discourse on AI in public safety and, more importantly, to make tangible improvements in the safety and well-being of the community.

\section{Related Works}\label{lit}

Ensuring public safety in urban environments is a complex challenge that varies significantly across different demographic groups and urban areas. Understanding these variations is crucial for developing effective security strategies. This literature review delves into existing research to explore public safety concerns and technology perceptions among diverse demographic groups.

\subsection{Diverse Public Safety Concerns}

Public safety concerns are not uniform and exhibit substantial variations across demographic groups. Vanolo \cite{vanolo2016smart} emphasizes the critical role of understanding citizen perspectives on safety within the context of smart cities. He underscores that citizen viewpoints are pivotal in shaping safety initiatives. Kummitha and Crutzen \cite{kummitha2017is} delve deeper into this topic, revealing how demographic factors such as age, gender, and socioeconomic status can significantly influence an individual's fear of crime and perceptions of safety.

Furthermore, research by Zandbergen and Uitermark \cite{zandbergen2019toward} underscores the significance of inclusivity in the smart city paradigm, particularly with regard to indigenous data. This suggests that factors like ethnicity and cultural background play substantial roles in shaping perceptions of safety and inclusion within urban environments.

\subsection{Perceptions of Video Surveillance Systems}

How individuals perceive video surveillance systems' effectiveness can vary widely depending on their demographic backgrounds. Shelton and Lodato \cite{shelton2019smart} argue that smart cities, which heavily rely on video surveillance, may inadvertently exacerbate urban inequalities. Different demographic groups may have disparate experiences with these surveillance systems, with concerns about privacy, security, and trust being shaped by age, income, and education levels.  There is considerable public concern about the tendency of facial recognition systems to misidentify minorities \cite{atay2021evaluation}. In 2020, the wrongful arrest of Robert Williams, an African American man in Detroit, due to a misidentification by a facial recognition system highlighted significant inaccuracies in these technologies, especially in identifying people of color. This incident catalyzed a broader debate on the racial biases inherent in facial recognition systems used by law enforcement \footnote{www.technologyreview.com}.

Angwin et al. \cite{angwin2016machine} shed light on the issue of algorithmic bias in predictive policing, which disproportionately affects certain demographic groups, especially communities of color. This suggests that demographics do not solely influence the perceptions of the effectiveness of video surveillance systems but are also shaped by concerns related to fairness and justice.

\subsection{Demographic Factors and AI-Driven Surveillance Technology}

Demographic background plays a substantial role in shaping perceptions of AI-driven video surveillance technology \cite{ho2022rethinking}. As digitalization and AI technologies become increasingly integrated into urban security, comprehending how different demographic groups view and interact with these technologies is paramount.

The literature on digitalization and smart cities, as discussed by Gabrys \cite{gabrys2014programming} and Cardullo and Kitchin \cite{cardullo2019smart}, indicates that these technologies impact demographic groups differently. Factors such as technological literacy, access to resources, and concerns about data privacy influence how individuals from diverse demographic backgrounds perceive and engage with AI-driven surveillance technology.

\section{Methodology}\label{Method}

The methodology employed for this research study involved in-person surveys conducted in Charlotte, North Carolina, during August and September 2023. This section outlines the data collection process, sampling methods, sample size, survey instrument, and the approach to data analysis.
 
\textbf{Data Collection:}
In-person surveys were the primary method of data collection for this study. The choice of in-person surveys was driven by the need to engage directly with the  Charlotte community, ensuring a more personal and comprehensive understanding of their public safety concerns. Our target population consisted of individuals likely to visit public places within the community. To gather a representative sample, we selected various community events in Charlotte, including Farmer's Markets, Church events, and regional library events.

\textbf{Sampling:}
Sampling methods involved a combination of convenience \cite{sedgwick2013convenience} and systematic sampling \cite{yates1948systematic}. Convenience sampling was used to select survey locations based on accessibility and population proportions. Systematic sampling was employed to ensure representation across various demographics, including age, gender, ethnicity, and geographic areas within Charlotte. For the selection of participants at each selected event, we approached every fifth person who entered and asked if they would participate in our study. This approach helped ensure that our sample was representative and avoided potential bias in participant selection.

\textbf{Sample Size:}
To determine our required sample size, we conducted a power analysis \cite{aberson2019applied} with a 95\% confidence level, a 5\% margin of error, an assumed population portion of 50\%, and a power rate 0.95. The analysis indicated we needed 385 responses to achieve the desired confidence level. We ultimately collected 410 responses to ensure a robust and diverse dataset.

\begin{figure}
    \centering
    \includegraphics[width=1\linewidth, trim=0 0 0 0, clip]{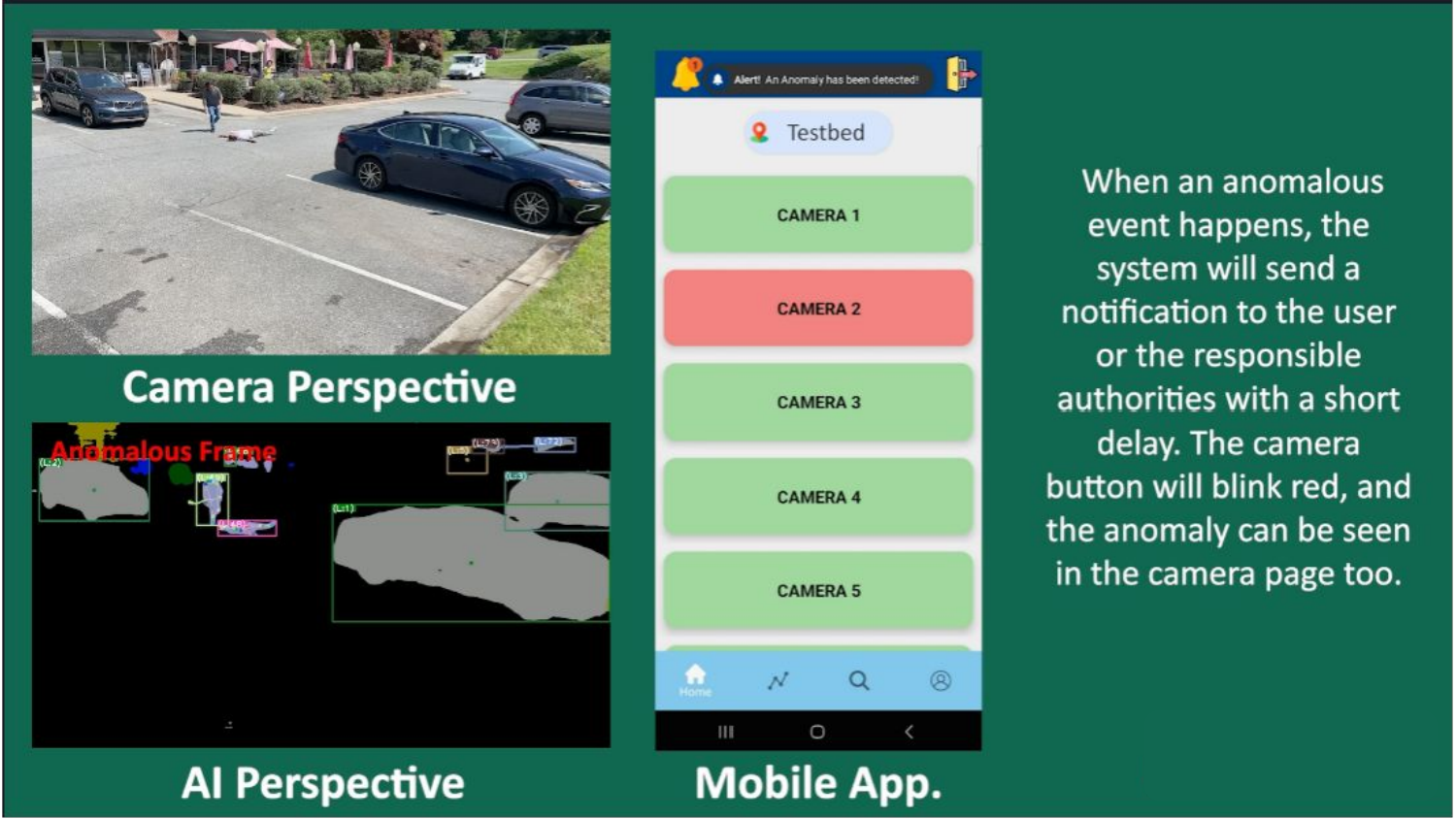}
    \caption{Screenshot of the demo video represented to the respondents. }\label{Demo}
\end{figure}

\textbf{Survey Instrument:}
The survey questionnaire was meticulously designed to collect data on various public safety concerns and the community's perceptions of existing and proposed AI-driven technologies. The questionnaire included sections on demographics, public safety concerns, prior experiences with surveillance technologies, and opinions on integrating AI solutions. To reduce the effect of respondents' different perceptions of AI-driven technologies, at the beginning of the last section, we asked people to watch a demo of AI-driven video surveillance coupled with an application that illustrates the system performance in case of anomaly occurrence. Figure \ref{Demo}, shows a screenshot of the 30-second video shown to the respondents. We utilized Ardabili et al. (2023) work to show an example of such a system \cite{ardabili2023understanding}.  

\textbf{Data Analysis:}
For the data analysis, we employed statistical techniques suitable for the dependent variables we worked with. When the dependent variables were binary (dummy) and ordinal, we used logistic regression \cite{lavalley2008logistic} and ordered logit models \cite{grilli2014ordered}, respectively. These methods allowed us to analyze the relationships between various demographic factors and the safety concerns, perceptions of video surveillance, and the acceptance of AI-driven technologies within our sample.

\textbf{Independet Variables:} In alignment with our research objectives, we utilized demographic factors as our independent variables. Specifically, we examined the respondents' age, gender, ethnicity, and educational background. For age, participants were prompted to select from the following brackets: "18-34", "35-49", "50-64", and "65 or older". It's worth noting that in compliance with our Institutional Review Board (IRB) protocol, we excluded individuals younger than 18. When inquiring about gender, respondents chose from "Male", "Female", or "Other". For ethnicity, the options included "White/Caucasian", "African American/Black", "Asian/Pacific Islander", "Hispanic/Latino", "Mixed", and "Other". Lastly, in terms of education, participants could select from "Less than High School", "High School Diploma", "Some College or Associate Degree", "Bachelor's Degree", or "Master's Degree or Higher". We ensured that respondents could abstain from answering any question if they chose. 

\textbf{Dependent Variables:} The subsequent survey sections constituted our dependent variables. In the second section, we inquired about the top three safety concerns of the respondents, the three public places where they felt most concerned about their safety, and their level of agreement with several statements using a five-point Likert scale, ranging from "Strongly Disagree" to "Strongly Agree". To initiate the third section, we provided an overview of current video surveillance technologies with the following description:
\begin{center}
"Understanding surveillance technologies: CCTV, license plate recognition, facial recognition, and
social media monitoring are used by law enforcement for public safety. These technologies aim to
prevent crime and assist with investigations but also raise privacy concerns. This survey
section explores your thoughts on these technologies and ways to address these concerns while
enhancing public safety."
\end{center}

Subsequently, we inquired about respondents' perceptions of the effectiveness of these technologies. Respondents were prompted to express their concern regarding certain specifications of these technologies using a five-point Likert scale, ranging from "Not Concerned at All" to "Extremely Concerned". We further asked them to convey their level of agreement with various statements related to the usage of these technologies and their feelings in the presence of video surveillance systems.

In the survey's concluding section, participants were first directed to view a thirty-second video showcasing an AI-driven smart video surveillance system, followed by related questions. Initially, they were asked to evaluate the benefits of such systems. This was succeeded by inquiries about their importance and apprehension on specific features of this technology. Additionally, they were posed with scenarios pertaining to their trust in the technology, given different thresholds of False Positives (FP) and False Negatives (FN). For instance, questions were structured as:

Would you trust this technology if:

\textit{False Negative:} 90\% of criminal activities were detected.

\textit{False Positive:} 50\% of alarms triggered were false.

To conclude, we sought their opinions on privacy concerns, especially considering the privacy-preserving measures embedded within this technology.
Table \ref{tabvar} represents the dependent variables, their types, and the associated coded version, which will be used in the rest of the paper. For clarity, not all dependent variables for safety concerns and areas are displayed in the table. A comprehensive discussion on these is available when discussing 'Data Preparation'.

\begin{table*}[]
\centering
\caption{Summary of dependent variables, their types, and what they measure.}\label{tabvar}
\begin{tabular}{|cccc|}
\hline
\rowcolor[HTML]{C0C0C0} 
\multicolumn{1}{|c|}{\cellcolor[HTML]{C0C0C0}\textbf{Questions}}                                                                                                                                                                         & \multicolumn{1}{c|}{\cellcolor[HTML]{C0C0C0}\textbf{Coded Version}} & \multicolumn{1}{c|}{\cellcolor[HTML]{C0C0C0}\textbf{Measurement}}                           & \textbf{Variable Type} \\ \hline
\rowcolor[HTML]{EFEFEF} 
\multicolumn{4}{|c|}{\cellcolor[HTML]{EFEFEF}\textbf{Section 2 (Genral Safety Concerns)}}                                                                                                                                                                                                                                                                                                                                             \\ \hline
\multicolumn{1}{|c|}{\begin{tabular}[c]{@{}c@{}}Select your three highest \\ public safety concerns:\end{tabular}}                                                                                                                       & \multicolumn{1}{c|}{Based on the concern type}                      & \multicolumn{1}{c|}{\cellcolor[HTML]{FFFFFF}0,1}                                            & Binary                 \\ \hline
\multicolumn{1}{|c|}{\begin{tabular}[c]{@{}c@{}}Select three local areas you are \\ most concern about your public safety\end{tabular}}                                                                                                  & \multicolumn{1}{c|}{Based on the area type}                         & \multicolumn{1}{c|}{0,1}                                                                    & Binary                 \\ \hline
\multicolumn{1}{|c|}{\begin{tabular}[c]{@{}c@{}}The presence of law enforcement \\ is sufficient to prevent crimes.\end{tabular}}                                                                                                        & \multicolumn{1}{c|}{Law\_Enforcement\_Presence}                     & \multicolumn{1}{c|}{\begin{tabular}[c]{@{}c@{}}Agreement \\ 5-Level Likert\end{tabular}}    & Ordinal                \\ \hline
\multicolumn{1}{|c|}{\begin{tabular}[c]{@{}c@{}}Law enforcement agencies effectively \\ address public safety issues.\end{tabular}}                                                                                                      & \multicolumn{1}{c|}{Effectively\_Address}                           & \multicolumn{1}{c|}{\begin{tabular}[c]{@{}c@{}}Agreement \\ 5-Level Likert\end{tabular}}    & Ordinal                \\ \hline
\multicolumn{1}{|c|}{\begin{tabular}[c]{@{}c@{}}Crime rates have increased in \\ my community in the last 10 years.\end{tabular}}                                                                                                        & \multicolumn{1}{c|}{Crime\_Increased}                               & \multicolumn{1}{c|}{\begin{tabular}[c]{@{}c@{}}Agreement \\ 5-Level Likert\end{tabular}}    & Ordinal                \\ \hline
\multicolumn{1}{|c|}{\begin{tabular}[c]{@{}c@{}}Law enforcement is adequately \\ protecting my personal information.\end{tabular}}                                                                                                       & \multicolumn{1}{c|}{Protecting\_Information}                        & \multicolumn{1}{c|}{\begin{tabular}[c]{@{}c@{}}Agreement \\ 5-Level Likert\end{tabular}}    & Ordinal                \\ \hline
\rowcolor[HTML]{EFEFEF} 
\multicolumn{4}{|c|}{\cellcolor[HTML]{EFEFEF}\textbf{Section 3 (Existing Technologies)}}                                                                                                                                                                                                                                                                                                                                              \\ \hline
\multicolumn{1}{|c|}{\begin{tabular}[c]{@{}c@{}}How Effective do you think video surveillance\\  technologies are in preventing crimes\end{tabular}}                                                                                     & \multicolumn{1}{c|}{Current\_Effectiveness}                         & \multicolumn{1}{c|}{\begin{tabular}[c]{@{}c@{}}Effectiveness\\ 5-Level Likert\end{tabular}} & Ordinal                \\ \hline
\multicolumn{1}{|c|}{Using facial recognition}                                                                                                                                                                                           & \multicolumn{1}{c|}{Current\_Facial\_Recognition}                   & \multicolumn{1}{c|}{\begin{tabular}[c]{@{}c@{}}Concern\\ 5-Level Likert\end{tabular}}       & Ordinal                \\ \hline
\multicolumn{1}{|c|}{Potential biases/discrimination}                                                                                                                                                                                    & \multicolumn{1}{c|}{Biases\_Discrimination}                         & \multicolumn{1}{c|}{\begin{tabular}[c]{@{}c@{}}Concern\\ 5-Level Likert\end{tabular}}       & Ordinal                \\ \hline
\multicolumn{1}{|c|}{Understanding how the data is being used}                                                                                                                                                                           & \multicolumn{1}{c|}{Data\_Usage}                                    & \multicolumn{1}{c|}{\begin{tabular}[c]{@{}c@{}}Concern\\ 5-Level Likert\end{tabular}}       & Ordinal                \\ \hline
\multicolumn{1}{|c|}{Understanding how the system works}                                                                                                                                                                                 & \multicolumn{1}{c|}{System\_Understanding}                          & \multicolumn{1}{c|}{\begin{tabular}[c]{@{}c@{}}Concern\\ 5-Level Likert\end{tabular}}       & Ordinal                \\ \hline
\multicolumn{1}{|c|}{Surveillance systems are a threat to privacy.}                                                                                                                                                                      & \multicolumn{1}{c|}{Privacy\_Threat}                                & \multicolumn{1}{c|}{\begin{tabular}[c]{@{}c@{}}Agreement \\ 5-Level Likert\end{tabular}}    & Ordinal                \\ \hline
\multicolumn{1}{|c|}{\begin{tabular}[c]{@{}c@{}}Video surveillance systems should \\ be used in law enforcement practices.\end{tabular}}                                                                                                 & \multicolumn{1}{c|}{Current\_Law-Use}                               & \multicolumn{1}{c|}{\begin{tabular}[c]{@{}c@{}}Agreement \\ 5-Level Likert\end{tabular}}    & Ordinal                \\ \hline
\multicolumn{1}{|c|}{\begin{tabular}[c]{@{}c@{}}The presence of video surveillance\\  systems makes me feel safer.\end{tabular}}                                                                                                         & \multicolumn{1}{c|}{Feel\_Safer}                                    & \multicolumn{1}{c|}{\begin{tabular}[c]{@{}c@{}}Agreement \\ 5-Level Likert\end{tabular}}    & Ordinal                \\ \hline
\rowcolor[HTML]{EFEFEF} 
\multicolumn{4}{|c|}{\cellcolor[HTML]{EFEFEF}\textbf{Section 4 (AI-driven Video Surveillance)}}                                                                                                                                                                                                                                                                                                                                       \\ \hline
\multicolumn{1}{|c|}{\begin{tabular}[c]{@{}c@{}}How beneficial do you think\\  this technology would be for public safety?\end{tabular}}                                                                                                 & \multicolumn{1}{c|}{Tech\_Benefcial}                                & \multicolumn{1}{c|}{\begin{tabular}[c]{@{}c@{}}Beneficial\\ 5-Level Likert\end{tabular}}    & Ordinal                \\ \hline
\multicolumn{1}{|c|}{Using AI for identifying criminal activities}                                                                                                                                                                       & \multicolumn{1}{c|}{Using\_AI}                                      & \multicolumn{1}{c|}{\begin{tabular}[c]{@{}c@{}}Importance\\ 5-Level Likert\end{tabular}}    & Ordinal                \\ \hline
\multicolumn{1}{|c|}{Not using facial recognition}                                                                                                                                                                                       & \multicolumn{1}{c|}{Tech\_Facial\_Recognition}                      & \multicolumn{1}{c|}{\begin{tabular}[c]{@{}c@{}}Importance\\ 5-Level Likert\end{tabular}}    & Ordinal                \\ \hline
\multicolumn{1}{|c|}{Use by law enforcement}                                                                                                                                                                                             & \multicolumn{1}{c|}{Tech\_Law\_Use}                                 & \multicolumn{1}{c|}{\begin{tabular}[c]{@{}c@{}}Importance\\ 5-Level Likert\end{tabular}}    & Ordinal                \\ \hline
\multicolumn{1}{|c|}{Citizens receiving real-time alerts}                                                                                                                                                                                & \multicolumn{1}{c|}{Notification}                                   & \multicolumn{1}{c|}{\begin{tabular}[c]{@{}c@{}}Importance\\ 5-Level Likert\end{tabular}}    & Ordinal                \\ \hline
\multicolumn{1}{|c|}{90\% of criminal activities are detected}                                                                                                                                                                           & \multicolumn{1}{c|}{90\_Detected}                                   & \multicolumn{1}{c|}{Yes, No}                                                                & Binary                 \\ \hline
\multicolumn{1}{|c|}{75\% of criminal activities are detected}                                                                                                                                                                           & \multicolumn{1}{c|}{75\_Detected}                                   & \multicolumn{1}{c|}{Yes, No}                                                                & Binary                 \\ \hline
\multicolumn{1}{|c|}{50\% of criminal activities are detected}                                                                                                                                                                           & \multicolumn{1}{c|}{50\_Detected}                                   & \multicolumn{1}{c|}{Yes, No}                                                                & Binary                 \\ \hline
\multicolumn{1}{|c|}{50\% of alarms raised are false}                                                                                                                                                                                    & \multicolumn{1}{c|}{50\_False\_Alarm}                               & \multicolumn{1}{c|}{Yes, No}                                                                & Binary                 \\ \hline
\multicolumn{1}{|c|}{25\% of alarms raised are false}                                                                                                                                                                                    & \multicolumn{1}{c|}{25\_False\_Alarm}                               & \multicolumn{1}{c|}{Yes, No}                                                                & Binary                 \\ \hline
\multicolumn{1}{|c|}{10\% of alarms raised are false}                                                                                                                                                                                    & \multicolumn{1}{c|}{10\_False\_Alarm}                               & \multicolumn{1}{c|}{Yes, No}                                                                & Binary                 \\ \hline
\multicolumn{1}{|c|}{\begin{tabular}[c]{@{}c@{}}Given that this technology is used to enhance \\ public safety without the use of facial recognition, \\ how concerned would you be about privacy \\ with this technology?\end{tabular}} & \multicolumn{1}{c|}{Privacy\_Concern}                               & \multicolumn{1}{c|}{\begin{tabular}[c]{@{}c@{}}Concern\\ 5-Level Likert\end{tabular}}       & Ordinal                \\ \hline
\end{tabular}
\end{table*}

\textbf{Data Preparation:}
In data preparation, each variable was adjusted to better represent the respondents' findings and ensure the results were presented clearly. For missing data, we replaced gaps with the most frequently occurring categories. For the "Age" variable, we combined the "65 or above" category with "50-64," labeling it "50 or above" to ensure more balanced age categories. Only one respondent selected "Other" for gender, so we grouped this response with the "Female" category, resulting in two gender categories. For "Ethnicity," the "Mixed" and "Other" categories were merged to achieve better data balance. Similarly, the "less than High School" and "High School Diploma" categories were merged and renamed "High School or Less."

It is important to note that the \textit{statsmodel} package in Python defaults to using the first group as the reference group\cite{mckinney2010data}. As a result, the "18-35" category serves as the reference group for "Age," "Male" is the reference for "Gender," "White/Caucasian" stands as the reference for "Ethnicity," and "High School or Less" is adopted as the reference group for "Education."

For questions related to "Safety Concerns" and "Area Apprehension", each choice was treated as a binary variable. For instance, the selection "Violent Crime" was deemed a dependent binary variable, assigned a value of "1" if chosen by the respondent and "0" otherwise. This approach yielded eight binary variables representing safety concerns and ten for public areas.

For other dependent variables on a five-level Likert scale, the initial choice was coded as "1". The second and third choices were consolidated and coded as "2", and the fourth and fifth were grouped and coded as "3". Additionally, the binary variables of FP and FN were coded as "0" for "No" and "1" for "Yes".

\textbf{Descriptive Statistics:}
Descriptive statistics were employed to summarize and understand the central tendencies and frequencies within the dataset. Table \ref{desc} represents the frequency count and the proportion of independent variables in each category. As mentioned in "\textit{Dependent Variable}", the data types in this study encompassed both binary and ordinal variables. More details of the statics on the dependent variables are represented in Subsection \ref{descriptive}. 
\begin{table*}
\centering
\caption{Descriptive Statistics for Independent Variables}\label{desc}
\begin{tabular}{llll}
\toprule
\textbf{Variable} & \textbf{Category/Group} & \textbf{Frequency Count} & \textbf{Proportion (\%)} \\
\midrule
\multirow{4}{*}{Age Ranges} & 18-34 & 189 & 46.67\% \\
& 35-49 & 117 & 28.89\% \\
& 50 or above & 99 & 24.44\% \\
\midrule
\multirow{2}{*}{Gender} & Female & 234 & 57.78\% \\
& Male & 171 & 42.22\% \\
\midrule
\multirow{5}{*}{Ethnicity} & White/Caucasian & 190 & 46.91\% \\
& African American/Black & 108 & 26.67\% \\
& Hispanic/Latino & 33 & 8.15\% \\
& Asian & 41 & 10.12\% \\
& Mixed and Other & 33 & 8.15\% \\
\midrule
\multirow{4}{*}{Educational Level} & Master's Degree or Higher & 122 & 30.12\% \\
& Bachelor's Degree & 155 & 38.27\% \\
& Some College or Associate Degree & 86 & 21.23\% \\
& High School or Less & 42 & 10.37\% \\
\bottomrule
\end{tabular} 
\end{table*}

\textbf{Quantitative Analysis:}

Quantitative data collected from individual respondents were subjected to various statistical analyses to derive meaningful insights into public safety concerns and community perceptions. The results of the statistical models provided valuable insights into the relationships between variables, offering a nuanced understanding of public safety concerns within the community. Significance tests, including p-values, were used to determine the robustness of the models and the statistical significance of the observed associations. The unit of analysis for this study was at the individual level.

\textbf{Logistic Regression:} Logistic regression is a statistical method suitable for analyzing the relationship between demographic factors and binary outcomes \cite{wright1995logistic}. Our study used it to assess safety concerns, perceptions of video surveillance, and the acceptance of AI-driven technologies. By estimating the probability of binary events, such as feeling safe or unsafe in public places, logistic regression helped us understand how demographic variables like age, gender, ethnicity, and socioeconomic status influence specific events' likelihood.
\begin{algorithm}
\caption{Logistic Regression (Statistical)}
\begin{algorithmic}
\State Initialize the model parameters: $\beta_0, \beta_1, \ldots, \beta_p$
\State Set the maximum number of iterations: $N$
\State Set the convergence threshold: $\epsilon$
\For{$i = 1$ to $N$}
    \State Compute the linear predictor: $\eta_i = \beta_0 + \beta_1 x_{i1} + \beta_2 x_{i2} + \ldots + \beta_p x_{ip}$
    \State Compute the predicted probabilities: $p_i = \frac{1}{1 + e^{-\eta_i}}$
    \State Compute the gradient of the log-likelihood: $\nabla L = \sum_{i=1}^{n} (y_i - p_i) \cdot \mathbf{X}_i$
    \State Update the model parameters: $\beta_0 \leftarrow \beta_0 + \alpha \cdot \nabla L_0, \beta_1 \leftarrow \beta_1 + \alpha \cdot \nabla L_1, \ldots, \beta_p \leftarrow \beta_p + \alpha \cdot \nabla L_p$
    \State Check for convergence: If $\|\nabla L\| < \epsilon$, break
\EndFor
\State \textbf{Return} the estimated model parameters: $\beta_0, \beta_1, \ldots, \beta_p$
\end{algorithmic}
\end{algorithm}

\textbf{Ordered Logit Model:} The ordered logit model is a variation of logistic regression designed for dependent variables with ordered categories, such as Likert scale responses \cite{seabold2010statsmodels}. Our study applied this model to explore the relationship between demographic factors and ordinal outcomes related to safety concerns, video surveillance perceptions, and AI technology acceptance. By considering the ordinal nature of responses, the ordered logit model allowed us to investigate how demographic variables influence transitions between ordered categories, providing insights into the impact of demographics on nuanced perceptions and acceptance levels.
\begin{algorithm}
\caption{Ordered Logit Model}
\begin{algorithmic}
\State Initialize the model parameters: $\beta_0, \beta_1, \ldots, \beta_p$
\State Set the maximum number of iterations: $N$
\State Set the convergence threshold: $\epsilon$
\For{$i = 1$ to $N$}
    \State Compute the linear predictors: $\eta_i = \beta_0 + \beta_1 x_{i1} + \beta_2 x_{i2} + \ldots + \beta_p x_{ip}$
    \State Compute the cut-points (thresholds): $\tau_1, \tau_2, \ldots, \tau_{J-1}$
    \State Determine the cumulative probabilities: $P(Y \leq j) = \frac{1}{1 + e^{-(\eta_i - \tau_j)}}$ for $j = 1, 2, \ldots, J-1$
    \State Compute the predicted category: $y_i = \min \{ j : \tau_j > \eta_i \}$
    \State Compute the gradient of the log-likelihood: $\nabla L = \sum_{i=1}^{n} (y_i - j_i) \cdot \mathbf{X}_i$
    \State Update the model parameters: $\beta_0 \leftarrow \beta_0 + \alpha \cdot \nabla L_0, \beta_1 \leftarrow \beta_1 + \alpha \cdot \nabla L_1, \ldots, \beta_p \leftarrow \beta_p + \alpha \cdot \nabla L_p$
    \State Check for convergence: If $\|\nabla L\| < \epsilon$, break
\EndFor
\State \textbf{Return} the estimated model parameters: $\beta_0, \beta_1, \ldots, \beta_p, \tau_1, \tau_2, \ldots, \tau_{J-1}$
\end{algorithmic}
\end{algorithm}

\textbf{Validity and Reliability:}
The survey employed in this study underwent rigorous validity and reliability assessments. To ensure content validity, the survey questions were checked and run by experts in the field, addressing various facets of public safety concerns and AI-driven technologies.

Reliability was assessed using Cronbach's Alpha ($\alpha$), which yielded a value of 0.84. This coefficient measures the internal consistency of the survey instrument \cite{amirrudin2021effect}, indicating that the questions within the survey reliably capture the constructs under investigation.

The survey's validity was enhanced through a pilot study involving a small sample of participants whose responses contributed to refining and improving the questionnaire. Additionally, using established scales, such as Likert scales, added construct validity to the survey instrument \cite{jebb2021review}.

\textbf{Ethical Considerations:}
The research study (IRB Approval Number: IRB-23-0944) adhered to rigorous ethical guidelines to protect participants' rights and privacy. Informed consent was obtained from all respondents, and participation was entirely voluntary. To uphold confidentiality, all data collected were anonymized and securely stored.

\section{Results}\label{results}
In this section, we present the findings of our research, which are divided into two subsections: Descriptive Analysis and Significance Analysis. The Descriptive Analysis provides a comprehensive overview of the participant's responses to the questions regardless of their demographic characteristics. The Significance Analysis delves into the statistical assessment of these demographic factors' impact on public attitudes and readiness for AI-driven safety measures.

\subsection{Descriptive Analysis}\label{descriptive}
In this section, we present an overarching view of survey responses, regardless of participants' demographic characteristics, and investigate the significance of correlations between dependent and independent variables.

Our analysis begins with a comprehensive exploration of responses to survey questions about public attitudes and readiness for AI-driven safety measures. By aggregating data from all demographic groups, we provide a holistic perspective on the overall sentiment within the surveyed population.

\begin{figure*}
    \centering
    \includegraphics[width=1\linewidth, trim=0 100 0 100, clip]{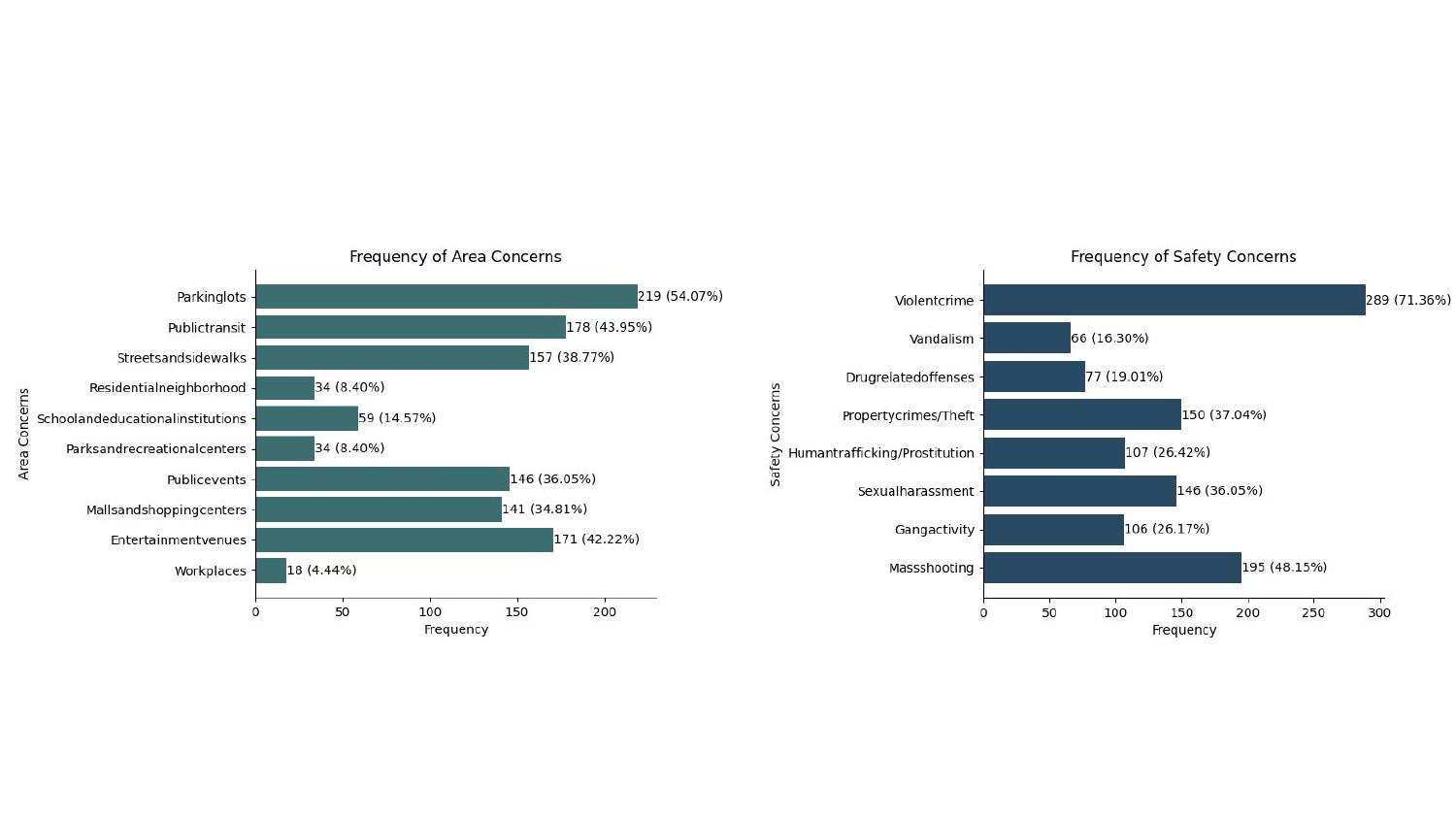}
    \caption{Distribution of Top Safety Concerns and High-Risk Areas Identified by Survey Respondents. The left panel shows the percentage of participants indicating areas where they feel most concerned for their safety, with parking lots being the primary area of unease. The right panel illustrates the prevalence of different safety concerns, with violent crime being the predominant worry among respondents.}\label{SConcern}
\end{figure*}
In figure \ref{SConcern}, we present the survey results where respondents were asked to identify their top three safety concerns and the locations where they feel most apprehensive regarding their safety. The data highlights the percentage of respondents who reported various safety concerns.

On the left, the "Frequency of Area Concerns" graph reveals that parking lots are the primary source of concern for the majority of participants, with 54.07\% indicating apprehension in these areas. This is closely followed by concerns in public transit at 43.95\% and entertainment venues at 42.22\%. In contrast, workplaces are the least cited areas of concern, with only 4.44\% of participants expressing unease.

Turning our attention to the right, the "Frequency of Safety Concerns" graph demonstrates that violent crime is overwhelmingly the most prominent worry for respondents, with 71.36\% selecting it as one of their top three concerns, followed by mass shootings with 48.15\%. Property crime and theft, with 37.04\%, and sexual harassment, at 36.05\%, demonstrate that these issues also weigh on the minds of community members. On the other end of the spectrum, vandalism is the least frequent concern, chosen by only 16.30\% of the surveyed individuals.

In essence, these results highlight that while specific locales such as parking lots and public transit are perceived as high-risk areas, concerns regarding violent crimes overshadow other safety issues across the board.

\begin{figure*}
    \centering
    \includegraphics[width=1\linewidth, trim=0 120 0 100, clip]{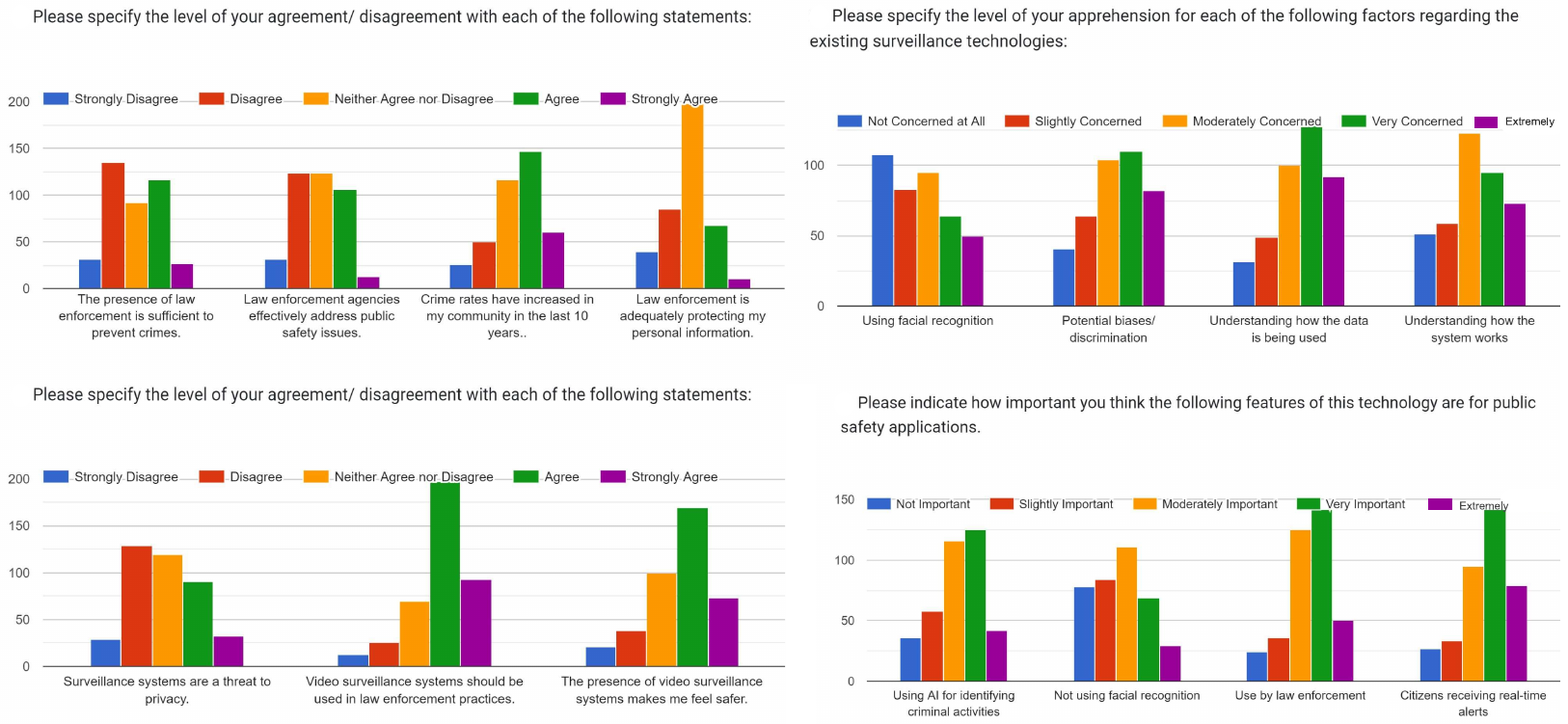}
    \caption{Distribution of responses to the questions with 5-Level Likert choices. }\label{Likert}
\end{figure*}

Figure \ref{Likert}, represents the distribution of responses to Five-Level Likert scale questions. In the top left graph, participants were asked about their agreement or disagreement level with certain law enforcement-related statements. The majority agreed or strongly agreed that the presence of law enforcement is sufficient to prevent crimes. However, responses were split on whether law enforcement agencies effectively address public safety issues. A significant number of respondents felt that crime rates have increased in their community over the last ten years. Lastly, most of the respondents chose neither agreement nor disagreement on whether the late enforcement protects their personal information. This finding underscores the inadequate sharing of information.

The top right graph captures public apprehension regarding various facets of existing surveillance technologies. Among the concerns, participants were extremely concerned about understanding how the system works, followed by understanding how the data is used. Most respondents expressed moderate concern about using facial recognition and its potential biases/discrimination. However, the leftward skew of the facial recognition graph, contrasted with the rightward skew for biases/discrimination, indicates a greater concern about biases and discrimination compared to using facial recognition technologies.  

The bottom left graph delves into public sentiment about surveillance technologies. A significant number of respondents felt that surveillance systems pose a threat to privacy. Despite these privacy concerns, there was a notable amount of support for using video surveillance systems in law enforcement practices. On the flip side, a majority also expressed that the presence of video surveillance systems made them feel safer.

The bottom right graph outlines the perceived importance of certain features of surveillance technology for public safety applications. The majority felt that using AI for identifying criminal activities and citizens receiving real-time alerts were extremely important. While the use of facial recognition was considered very important by some, a notable group also found it not important. The use by law enforcement was seen as moderately to very important by most respondents.

\begin{figure}
    \centering
    \includegraphics[width=1\linewidth, trim=0 250 0 180, clip]{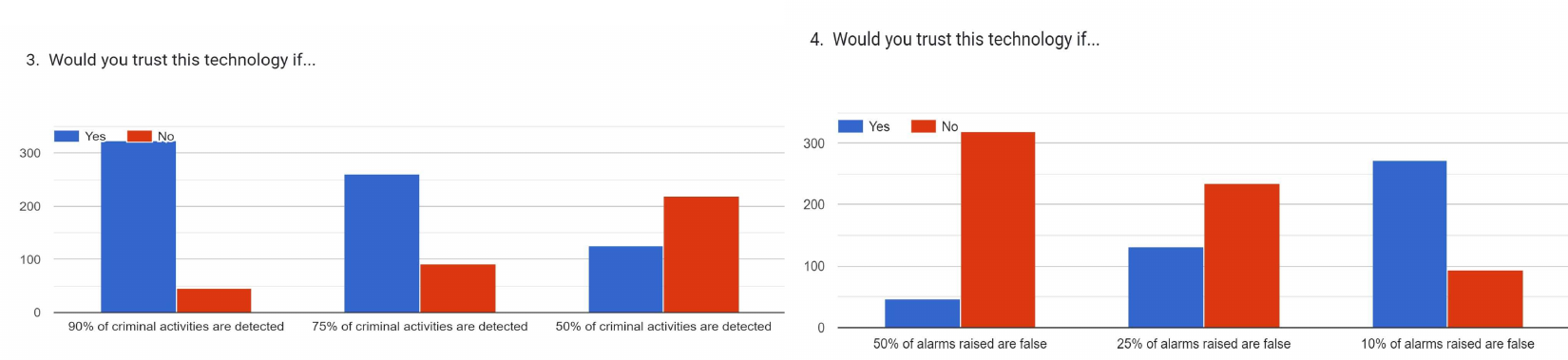}
    \caption{Distribution of responses to False-Negative and False-Positive Questions. }\label{FPFN}
\end{figure}

Figure \ref{FPFN}, illustrates the responses to False-Negative and False-Positive questions. In the left graph, respondents were asked if they would trust a technology based on its capability to detect criminal activities. The graph illustrates a clear correlation between the efficiency of the technology and the trust placed in it. Nearly all participants would trust the technology if it could detect 90\% of criminal activities. Trust diminishes as the detection rate drops, with a significant drop in trust when the technology can only detect 50\% of criminal activities.

The right graph assesses trust in the technology based on its false alarm rate. It's evident that as the percentage of false alarms increases, trust in the technology decreases. While a majority would still trust the technology if 10\% of alarms raised were false, this trust significantly wanes when half the alarms raised are false. Interestingly, even at a 25\% false alarm rate, respondents are split between trusting and not trusting the technology.

\subsection{Significance Analysis}\label{Sig}
The Significance Analysis delves deeper into the data, employing statistical methods to examine the relationships and significance of various demographic factors in shaping public attitudes toward AI-integrated safety measures. Based on the type of dependent variables, We utilize methods like logistic regression for binary dependent variables and ordered logistic regression for ordinal dependent variables to assess the impact of age, gender, ethnicity, and educational level on survey responses. These analyses unveil the key factors influencing public perceptions and preferences regarding integrating AI technology into urban safety initiatives.
The logistic regression analysis offers valuable insights into the relationship between our demographic factors and binary dependent variables. Table \ref{LRT} presents the results from logistic regression analyses on binary variables. Each row represents a different dependent variable, the outcomes we're interested in. The columns represent the independent variables or predictors that may influence the outcomes. In each table cell, the top number represents the coefficients, and the other numbers in parentheses indicate the p-values. The p-values lower than 0.05 are marked with stars. In the next paragraphs, we will discuss valuable insights in Table \ref{LRT}. 

 "Age" is an important factor regarding the safety concerns. As we can see in Table \ref{LRT}, "Age" is positively associated with "Violent Crime," "Vandalism," and "Gang Activity." This relationship suggests that holding all other variables constant, older people are more likely to consider "Violent Crime," "Vandalism," and "Gang Activity" as a source of concern compared to the reference group. More specifically, the probability of older groups selecting the mentioned safety concerns will increase by 1.37, 1.44, and 1.51 percentage points, respectively, compared to the reference group. On the other hand, there is a negative association between the "Age" variable and "Human Trafficking/Prostitution" as well as "Sexual Harassment." This suggests that the probability of selecting these safety concerns decreases among older people compared to younger ones.  

\textbf{Insight 1- Age-Related Concerns Reflect Societal Challenges}: \textit{There's a distinct divide in concerns based on "\textbf{Age}". Older individuals are more worried about what can be termed as "Turbulent Activities" (like Violent crimes, Vandalism, and gang activities), which can be interpreted as disturbances that directly affect community peace. In contrast, younger people are more concerned about "Social Improprieties" like human trafficking and sexual harassment. This possibly indicates that younger generations are more attuned to covert societal issues that might not be as visible but have deeply personal and long-term implications.} 

The results in the table show that compared to "White/Caucasian" group, minority ethnic groups like "Asians," "Latinos," and "Other" are more likely to consider "Vandalism" and "Drug Related Offenses" as their primary source of concern. More precisely, holding other variables constant, in comparison to the reference group, the log odds of selecting these choices increased among the mentioned minority groups by 1.3, 0.96, and 1.07, respectively. Specifically, taking a closer look at "Other" reveals that this group of respondents is also more likely to feel unsafe in there "Neighborhood" compared to the reference group.   

\textbf{Insight 2- Ethnic-Related Safety Concerns}: \textit{Minority ethnic groups, notably "\textbf{Asian}," "\textbf{Latinos}," and "\textbf{Other}," manifest heightened concerns about disruptions like "Vandalism" and "Drug Related Offences." Their elevated apprehensions might stem from various factors, including cultural perceptions, historical experiences, or specific community challenges, underscoring the need for targeted safety measures and community outreach for these particular groups \cite{kirmayer2009community, yoshihama2000reinterpreting, giles2007accommodating, wu2011race}.}

The coefficients of "African Americans" for "Parking Lots" and "Streets and Sidewalks" are negative, suggesting a negative association between this group and the aforementioned places compared to "White" individuals. On the other hand, the positive coefficient for the "Malls and Shopping Centers" illustrates that African Americans are more likely to feel unsafe at these places. 

\textbf{Insight 3- Different Area Perceptions among Ethnic Groups}: \textit{There's a distinct spatial pattern in safety perceptions among "\textbf{African Americans}" compared to "White" individuals. They tend to feel more vulnerable in enclosed spaces like "Malls" than in open areas like "Parking Lots." This divergence from the reference group may be rooted in historical, social, or individual experiences \cite{logan2011separate}. Recognizing these nuanced feelings of safety is essential for crafting more inclusive security measures and public policies.}

The positive coefficients of "Age" and "Gender" for the "Tech-Beneficial" variable in Table \ref{tbl2} suggest a positive significant association between these independent variables and the dependent variable. More precisely, the 0.1722 coefficient for the "Age" variable indicates that holding other variables constant, older people are more likely to find the technology more beneficial. Also, being a "Female" increased the log odds of considering the technology more beneficial by 0.386 compared to Males. This finding is consistent with the other findings in the table. For example, "Females" are less likely to believe that surveillance technology is a "Threat to Privacy" than males. Also, "Females" indicate more "Agreement" to feel safe in the presence of video surveillance while they are less concerned with the use of "Facial Technology." 

\textbf{Insight 4- Demographic Influences on Perception of Technology}: \textit{The perception and acceptance of the new technology vary significantly based on "\textbf{Gender}" and "\textbf{Age}." 
While females generally view facial recognition technology as a source of safety and express fewer concerns about its impact on privacy, older individuals, on the other hand, particularly appreciate its potential for aiding law enforcement. This suggests that different demographic groups prioritize the benefits and risks of new technologies differently, with females valuing the security it offers and older individuals seeing its potential as a tool for established agencies.}

False Positives (FP) and False Negatives (FN) are two important parameters in assessing the performance of an AI-enabled system \cite{noghre2023understanding}. In public safety, some stakeholders, such as technology developers, are more interested in minimizing the FN. At the same time, the end-users might be more interested in receiving fewer false alarms, i.e., minimizing FPs. Our findings show that "Education" is negatively associated with accepting 50\% False Negative (FN) rate. More accurately, holding higher degrees decreases the probability of accepting this False Negative rate by 1.26 percentage points. Accepting 50\%  FN means trusting the technology whose performance is slightly better than flipping a coin to detect criminal activities. On the other hand, minority ethnic groups such as African Americans and Asians indicate a significant positive association with accepting 50\% FP. It means that the people most likely to be misidentified by this technology are most tolerant of being misidentified. 

\textbf{Insight 5- Socioeconomic Factors Influence Concerns}: \textit{"\textbf{Education} plays a pivotal role in shaping concerns. Individuals with higher education seem less concerned about human trafficking, suggesting that a mediator explainable variable such as income might be a protective factor against such concerns \cite{jac2012determinants}. Additionally, the relationship between \textbf{Education} and the acceptable rate of false negatives suggests that those with more education might have higher expectations for the technology's accuracy. Meanwhile, \textbf{minority groups} are more tolerant of higher false positive rates than white individuals, hinting at different thresholds of technological trust or potential discrepancies in their experiences with security technologies.}

\textbf{Insight 6- Universal Concern over Biases in Technology}: \textit{The fear and concern surrounding "\textbf{Biases and Discrimination} in technology cut across all demographics, illustrating its universal importance. Remarkably, none of the demographic variables stood out as significant in the data. This suggests that irrespective of age, gender, ethnicity, or education, the dread of biases is a shared sentiment. It underscores the necessity for technology developers and policymakers to ensure fairness and address potential biases proactively.}
\begin{table*}[]
\centering
\caption{Summary of Coefficients and P-values for Logistic Regression Analysis on Binary Variables. P-values are shown in parentheses, and the significance at 0.95 level is marked by *. }\label{LRT}
\begin{tabular}{ccllllll}
\toprule
\begin{tabular}[c]{@{}c@{}}\textbf{Dependent}\\ \textbf{ Variables}\end{tabular} &  \textbf{Age} & \textbf{Gender} & \textbf{Education} & \begin{tabular}[c]{@{}c@{}}\textbf{African}\\ \textbf{ American}\end{tabular} & \textbf{Asian} & \textbf{ Hispanic} & \textbf{Other} \\ \hline\hline

ViolentCrime  & \begin{tabular}[c]{@{}c@{}}0.3189\\ (0.029)*\end{tabular} & \begin{tabular}[c]{@{}c@{}}-0.1489\\ (0.516)\end{tabular} & \begin{tabular}[c]{@{}c@{}}0.0570\\ (0.638)\end{tabular} & \begin{tabular}[c]{@{}c@{}}0.5254\\ (0.076)\end{tabular} & \begin{tabular}[c]{@{}c@{}}-0.1967\\ (0.590)\end{tabular} & \begin{tabular}[c]{@{}c@{}}0.0267\\ (0.948)\end{tabular} & \begin{tabular}[c]{@{}c@{}}0.5233\\ (0.252)\end{tabular} \\ \hline
Vandalism  & \begin{tabular}[c]{@{}c@{}}0.3626\\ (0.032)*\end{tabular} & \begin{tabular}[c]{@{}c@{}}0.0884\\ (0.753)\end{tabular} & \begin{tabular}[c]{@{}c@{}}0.1327\\ (0.369)\end{tabular} & \begin{tabular}[c]{@{}c@{}}0.0337\\ (0.923)\end{tabular} & \begin{tabular}[c]{@{}c@{}}0.2392\\ (0.613)\end{tabular} & \begin{tabular}[c]{@{}c@{}}0.9635\\ (0.036)*\end{tabular} & \begin{tabular}[c]{@{}c@{}}0.4614\\ (0.338)\end{tabular} \\ \hline
DrugRelatedOffenses  & \begin{tabular}[c]{@{}c@{}}0.2561\\ (0.116)\end{tabular} & \begin{tabular}[c]{@{}c@{}}-0.1933\\ (0.464)\end{tabular} & \begin{tabular}[c]{@{}c@{}}0.0126\\ (0.928)\end{tabular} & \begin{tabular}[c]{@{}c@{}}-0.0228\\ (0.948)\end{tabular} & \begin{tabular}[c]{@{}c@{}}1.2856\\ (0.001)*\end{tabular} & \begin{tabular}[c]{@{}c@{}}0.7335\\ (0.116)\end{tabular} & \begin{tabular}[c]{@{}c@{}}1.0740\\ (0.012)*\end{tabular} \\ \hline
PropertyCrimes/Theft  & \begin{tabular}[c]{@{}c@{}}-0.2192\\ (0.099)\end{tabular} & \begin{tabular}[c]{@{}c@{}}-0.2751\\ (0.195)\end{tabular} & \begin{tabular}[c]{@{}c@{}}0.0398\\ (0.723)\end{tabular} & \begin{tabular}[c]{@{}c@{}}0.0908\\ (0.725)\end{tabular} & \begin{tabular}[c]{@{}c@{}}-0.8778\\ (0.032)*\end{tabular} & \begin{tabular}[c]{@{}c@{}}-0.5290\\ (0.214)\end{tabular} & \begin{tabular}[c]{@{}c@{}}0.1495\\ (0.698)\end{tabular} \\ \hline
Human Trafficking/Prostitution  & \begin{tabular}[c]{@{}c@{}}-0.4305\\ (0.006)*\end{tabular} & \begin{tabular}[c]{@{}c@{}}0.3514\\ (0.145)\end{tabular} & \begin{tabular}[c]{@{}c@{}}-0.3602\\ (0.004)*\end{tabular} & \begin{tabular}[c]{@{}c@{}}0.0839\\ (0.771)\end{tabular} & \begin{tabular}[c]{@{}c@{}}-0.5149\\ (0.259)\end{tabular} & \begin{tabular}[c]{@{}c@{}}0.1647\\ (0.692)\end{tabular} & \begin{tabular}[c]{@{}c@{}}-0.2911\\ (0.536)\end{tabular} \\ \hline
Sexual Harassment  & \begin{tabular}[c]{@{}c@{}}-0.9672\\ (0.000)*\end{tabular} & \begin{tabular}[c]{@{}c@{}}1.5283\\ (0.000)*\end{tabular} & \begin{tabular}[c]{@{}c@{}}0.2200\\ (0.087)\end{tabular} & \begin{tabular}[c]{@{}c@{}}-0.1792\\ (0.548)\end{tabular} & \begin{tabular}[c]{@{}c@{}}0.4100\\ (0.291)\end{tabular} & \begin{tabular}[c]{@{}c@{}}0.0990\\ (0.817)\end{tabular} & \begin{tabular}[c]{@{}c@{}}0.1421\\ (0.744)\end{tabular} \\ \hline
Gang Activity  & \begin{tabular}[c]{@{}c@{}}0.4135\\ (0.004)*\end{tabular} & \begin{tabular}[c]{@{}c@{}}-0.2068\\ (0.375)\end{tabular} & \begin{tabular}[c]{@{}c@{}}0.1306\\ (0.296)\end{tabular} & \begin{tabular}[c]{@{}c@{}}-0.3963\\ (0.175)\end{tabular} & \begin{tabular}[c]{@{}c@{}}-0.0876\\ (0.825)\end{tabular} & \begin{tabular}[c]{@{}c@{}}-0.0741\\ (0.869)\end{tabular} & \begin{tabular}[c]{@{}c@{}}-0.3021\\ (0.498)\end{tabular} \\ \hline
Mass Shooting  & \begin{tabular}[c]{@{}c@{}}-0.0184\\ (0.887)\end{tabular} & \begin{tabular}[c]{@{}c@{}}-0.2440\\ (0.243)\end{tabular} & \begin{tabular}[c]{@{}c@{}}-0.0365\\ (0.738)\end{tabular} & \begin{tabular}[c]{@{}c@{}}0.0939\\ (0.709)\end{tabular} & \begin{tabular}[c]{@{}c@{}}-0.6493\\ (0.070)\end{tabular} & \begin{tabular}[c]{@{}c@{}}-0.4099\\ (0.291)\end{tabular} & \begin{tabular}[c]{@{}c@{}}-1.6255\\ (0.001)*\end{tabular} \\ \hline
Parking Lots  & \begin{tabular}[c]{@{}c@{}}0.0670\\ (0.602)\end{tabular} & \begin{tabular}[c]{@{}c@{}}0.3442\\ (0.096)\end{tabular} & \begin{tabular}[c]{@{}c@{}}0.0518\\ (0.635)\end{tabular} & \begin{tabular}[c]{@{}c@{}}-0.6116\\ (0.016)*\end{tabular} & \begin{tabular}[c]{@{}c@{}}-0.4100\\ (0.241)\end{tabular} & \begin{tabular}[c]{@{}c@{}}0.4887\\ (0.238)\end{tabular} & \begin{tabular}[c]{@{}c@{}}-0.1596\\ (0.676)\end{tabular} \\ \hline
Public Transit & \begin{tabular}[c]{@{}c@{}}-0.3848\\ (0.003)*\end{tabular} & \begin{tabular}[c]{@{}c@{}}0.1449\\ (0.487)\end{tabular} & \begin{tabular}[c]{@{}c@{}}-0.0491\\ (0.653)\end{tabular} & \begin{tabular}[c]{@{}c@{}}0.0038\\ (0.988)\end{tabular} & \begin{tabular}[c]{@{}c@{}}-0.1046\\ (0.768)\end{tabular} & \begin{tabular}[c]{@{}c@{}}0.1116\\ (0.774)\end{tabular} & \begin{tabular}[c]{@{}c@{}}0.5144\\ (0.183)\end{tabular} \\ \hline
Streets and Sidewalks  & \begin{tabular}[c]{@{}c@{}}-0.2371\\ (0.073)\end{tabular} & \begin{tabular}[c]{@{}c@{}}-0.1145\\ (0.588)\end{tabular} & \begin{tabular}[c]{@{}c@{}}0.0207\\ (0.854)\end{tabular} & \begin{tabular}[c]{@{}c@{}}-0.5493\\ (0.043)*\end{tabular} & \begin{tabular}[c]{@{}c@{}}0.3792\\ (0.278)\end{tabular} & \begin{tabular}[c]{@{}c@{}}0.0456\\ (0.907)\end{tabular} & \begin{tabular}[c]{@{}c@{}}0.0815\\ (0.832)\end{tabular} \\ \hline
Residential Neighborhood  & \begin{tabular}[c]{@{}c@{}}-0.1085\\ (0.644)\end{tabular} & \begin{tabular}[c]{@{}c@{}}0.4661\\ (0.229)\end{tabular} & \begin{tabular}[c]{@{}c@{}}0.0465\\ (0.816)\end{tabular} & \begin{tabular}[c]{@{}c@{}}0.2907\\ (0.540)\end{tabular} & \begin{tabular}[c]{@{}c@{}}0.6731\\ (0.236)\end{tabular} & \begin{tabular}[c]{@{}c@{}}-0.1082\\ (0.892)\end{tabular} & \begin{tabular}[c]{@{}c@{}}1.2079\\ (0.027)*\end{tabular} \\ \hline
School and Educational Institutions & \begin{tabular}[c]{@{}c@{}}0.1689\\ (0.342)\end{tabular} & \begin{tabular}[c]{@{}c@{}}-0.3529\\ (0.221)\end{tabular} & \begin{tabular}[c]{@{}c@{}}-0.2550\\ (0.085)\end{tabular} & \begin{tabular}[c]{@{}c@{}}0.1126\\ (0.755)\end{tabular} & \begin{tabular}[c]{@{}c@{}}0.1024\\ (0.847)\end{tabular} & \begin{tabular}[c]{@{}c@{}}0.5794\\ (0.240)\end{tabular} & \begin{tabular}[c]{@{}c@{}}0.6232\\ (0.196)\end{tabular} \\ \hline
Parks and Recreational Centers  & \begin{tabular}[c]{@{}c@{}}0.1630\\ (0.470)\end{tabular} & \begin{tabular}[c]{@{}c@{}}-1.0176\\ (0.008)*\end{tabular} & \begin{tabular}[c]{@{}c@{}}0.2987\\ (0.158)\end{tabular} & \begin{tabular}[c]{@{}c@{}}-0.7630\\ (0.195)\end{tabular} & \begin{tabular}[c]{@{}c@{}}0.3711\\ (0.509)\end{tabular} & \begin{tabular}[c]{@{}c@{}}0.7681\\ (0.214)\end{tabular} & \begin{tabular}[c]{@{}c@{}}0.6578\\ (0.245)\end{tabular} \\ \hline
Public Events  & \begin{tabular}[c]{@{}c@{}}-0.0302\\ (0.819)\end{tabular} & \begin{tabular}[c]{@{}c@{}}0.0184\\ (0.931)\end{tabular} & \begin{tabular}[c]{@{}c@{}}0.0405\\ (0.719)\end{tabular} & \begin{tabular}[c]{@{}c@{}}0.3882\\ (0.130)\end{tabular} & \begin{tabular}[c]{@{}c@{}}-0.1548\\ (0.677)\end{tabular} & \begin{tabular}[c]{@{}c@{}}-0.6670\\ (0.145)\end{tabular} & \begin{tabular}[c]{@{}c@{}}0.2050\\ (0.598)\end{tabular} \\ \hline
Malls and Shopping Centers  & \begin{tabular}[c]{@{}c@{}}0.2375\\ (0.076)\end{tabular} & \begin{tabular}[c]{@{}c@{}}0.0907\\ (0.679)\end{tabular} & \begin{tabular}[c]{@{}c@{}}-0.0517\\ (0.652)\end{tabular} & \begin{tabular}[c]{@{}c@{}}0.5747\\ (0.025)*\end{tabular} & \begin{tabular}[c]{@{}c@{}}-0.0134\\ (0.971)\end{tabular} & \begin{tabular}[c]{@{}c@{}}-1.0223\\ (0.048)*\end{tabular} & \begin{tabular}[c]{@{}c@{}}-0.4629\\ (0.290)\end{tabular} \\ \hline
Entertainment Venues  & \begin{tabular}[c]{@{}c@{}}-0.1214\\ (0.346)\end{tabular} & \begin{tabular}[c]{@{}c@{}}0.1932\\ (0.353)\end{tabular} & \begin{tabular}[c]{@{}c@{}}0.1223\\ (0.264)\end{tabular} & \begin{tabular}[c]{@{}c@{}}-0.0830\\ (0.742)\end{tabular} & \begin{tabular}[c]{@{}c@{}}-0.4492\\ (0.211)\end{tabular} & \begin{tabular}[c]{@{}c@{}}-0.3761\\ (0.344)\end{tabular} & \begin{tabular}[c]{@{}c@{}}-0.5098\\ (0.201)\end{tabular} \\ \hline
Workplaces  & \begin{tabular}[c]{@{}c@{}}-0.0369\\ (0.905)\end{tabular} & \begin{tabular}[c]{@{}c@{}}0.7149\\ (0.187)\end{tabular} & \begin{tabular}[c]{@{}c@{}}-0.1116\\ (0.667)\end{tabular} & \begin{tabular}[c]{@{}c@{}}0.0661\\ (0.907)\end{tabular} & \begin{tabular}[c]{@{}c@{}}0.0180\\ (0.982)\end{tabular} & \begin{tabular}[c]{@{}c@{}}-38.6322\\ (1.000)\end{tabular} & \begin{tabular}[c]{@{}c@{}}-0.4878\\ (0.650)\end{tabular} \\ \hline
90\_Detected  & \begin{tabular}[c]{@{}c@{}}-0.1774\\ (0.377)\end{tabular} & \begin{tabular}[c]{@{}c@{}}0.3267\\ (0.325)\end{tabular} & \begin{tabular}[c]{@{}c@{}}0.1314\\ (0.437)\end{tabular} & \begin{tabular}[c]{@{}c@{}}-0.3994\\ (0.291)\end{tabular} & \begin{tabular}[c]{@{}c@{}}32.2047\\ (1.000)\end{tabular} & \begin{tabular}[c]{@{}c@{}}0.0928\\ (0.889)\end{tabular} & \begin{tabular}[c]{@{}c@{}}-0.4624\\ (0.396)\end{tabular} \\ \hline
75\_Detected  & \begin{tabular}[c]{@{}c@{}}0.1742\\ (0.264)\end{tabular} & \begin{tabular}[c]{@{}c@{}}0.4531\\ (0.064)\end{tabular} & \begin{tabular}[c]{@{}c@{}}-0.0690\\ (0.598)\end{tabular} & \begin{tabular}[c]{@{}c@{}}-0.0175\\ (0.955)\end{tabular} & \begin{tabular}[c]{@{}c@{}}0.0425\\ (0.920)\end{tabular} & \begin{tabular}[c]{@{}c@{}}-0.3568\\ (0.416)\end{tabular} & \begin{tabular}[c]{@{}c@{}}0.2324\\ (0.633)\end{tabular} \\ \hline
50\_Detected  & \begin{tabular}[c]{@{}c@{}}0.1598\\ (0.244)\end{tabular} & \begin{tabular}[c]{@{}c@{}}0.3283\\ (0.147)\end{tabular} & \begin{tabular}[c]{@{}c@{}}-0.2286\\ (0.049)*\end{tabular} & \begin{tabular}[c]{@{}c@{}}0.1633\\ (0.538)\end{tabular} & \begin{tabular}[c]{@{}c@{}}-0.0855\\ (0.827)\end{tabular} & \begin{tabular}[c]{@{}c@{}}-0.6354\\ (0.172)\end{tabular} & \begin{tabular}[c]{@{}c@{}}-0.1726\\ (0.685)\end{tabular} \\ \hline
50\_False\_Alarm  & \begin{tabular}[c]{@{}c@{}}0.2508\\ (0.222)\end{tabular} & \begin{tabular}[c]{@{}c@{}}-0.3429\\ (0.296)\end{tabular} & \begin{tabular}[c]{@{}c@{}}-0.3230\\ (0.056)\end{tabular} & \begin{tabular}[c]{@{}c@{}}1.0075\\ (0.017)*\end{tabular} & \begin{tabular}[c]{@{}c@{}}1.9973\\ (0.000)*\end{tabular} & \begin{tabular}[c]{@{}c@{}}-0.0269\\ (0.973)\end{tabular} & \begin{tabular}[c]{@{}c@{}}1.0729\\ (0.065)\end{tabular} \\ \hline
25\_False\_Alarm  & \begin{tabular}[c]{@{}c@{}}-0.1885\\ (0.168)\end{tabular} & \begin{tabular}[c]{@{}c@{}}-0.1151\\ (0.598)\end{tabular} & \begin{tabular}[c]{@{}c@{}}0.0293\\ (0.801)\end{tabular} & \begin{tabular}[c]{@{}c@{}}-0.2270\\ (0.402)\end{tabular} & \begin{tabular}[c]{@{}c@{}}-0.1964\\ (0.598)\end{tabular} & \begin{tabular}[c]{@{}c@{}}-0.9033\\ (0.061)\end{tabular} & \begin{tabular}[c]{@{}c@{}}0.1924\\ (0.621)\end{tabular} \\ \hline
10\_False\_Alarm  & \begin{tabular}[c]{@{}c@{}}-0.1448\\ (0.344)\end{tabular} & \begin{tabular}[c]{@{}c@{}}0.3155\\ (0.202)\end{tabular} & \begin{tabular}[c]{@{}c@{}}0.1046\\ (0.413)\end{tabular} & \begin{tabular}[c]{@{}c@{}}-0.7355\\ (0.015)*\end{tabular} & \begin{tabular}[c]{@{}c@{}}-0.7490\\ (0.068)\end{tabular} & \begin{tabular}[c]{@{}c@{}}-0.5579\\ (0.227)\end{tabular} & \begin{tabular}[c]{@{}c@{}}-0.8343\\ (0.053)\end{tabular} \\ \bottomrule
\end{tabular}
\end{table*}

\begin{table*}[]
\centering
\caption{Summary of Coefficients and P-values for Ordered Logit Analysis on Ordinal Variables. P-values are shown in parentheses, and the significance at 0.95 level is marked by *. }\label{tbl2}
\begin{tabular}{ccllllll}
\toprule
\begin{tabular}[c]{@{}c@{}}\textbf{Dependent}\\ \textbf{ Variables}\end{tabular} &  \textbf{Age} & \textbf{Gender} & \textbf{Education} & \begin{tabular}[c]{@{}c@{}}\textbf{African}\\ \textbf{ American}\end{tabular} & \textbf{Asian} & \textbf{ Hispanic} & \textbf{Other} \\ \hline\hline
Law\_Enforcement\_Presence & \begin{tabular}[c]{@{}c@{}}-0.0973\\ (0.183)\end{tabular} & \begin{tabular}[c]{@{}c@{}}-0.2018\\ (0.086)*\end{tabular} & \begin{tabular}[c]{@{}c@{}}-0.0553\\ (0.371)\end{tabular} & \begin{tabular}[c]{@{}c@{}}0.1246\\ (0.387)\end{tabular} & \begin{tabular}[c]{@{}c@{}}0.0277\\ (0.887)\end{tabular} & \begin{tabular}[c]{@{}c@{}}-0.0258\\ (0.906)\end{tabular} & \begin{tabular}[c]{@{}c@{}}0.1856\\ (0.397)\end{tabular} \\ \hline
Effectively\_Address & \begin{tabular}[c]{@{}c@{}}0.1591\\ (0.025)*\end{tabular} & \begin{tabular}[c]{@{}c@{}}-0.1044\\ (0.366)\end{tabular} & \begin{tabular}[c]{@{}c@{}}-0.0732\\ (0.228)\end{tabular} & \begin{tabular}[c]{@{}c@{}}-0.1146\\ (0.421)\end{tabular} & \begin{tabular}[c]{@{}c@{}}0.2480\\ (0.201)\end{tabular} & \begin{tabular}[c]{@{}c@{}}-0.1131\\ (0.599)\end{tabular} & \begin{tabular}[c]{@{}c@{}}0.2326\\ (0.278)\end{tabular} \\ \hline
Crime\_Increased & \begin{tabular}[c]{@{}c@{}}0.0811\\ (0.268)\end{tabular} & \begin{tabular}[c]{@{}c@{}}0.1950\\ (0.097)\end{tabular} & \begin{tabular}[c]{@{}c@{}}0.0644\\ (0.301)\end{tabular} & \begin{tabular}[c]{@{}c@{}}-0.0177\\ (0.904)\end{tabular} & \begin{tabular}[c]{@{}c@{}}-0.1166\\ (0.558)\end{tabular} & \begin{tabular}[c]{@{}c@{}}0.0567\\ (0.800)\end{tabular} & \begin{tabular}[c]{@{}c@{}}-0.1249\\ (0.570)\end{tabular} \\ \hline
Protecting\_Information & \begin{tabular}[c]{@{}c@{}}0.1553\\ (0.027)*\end{tabular} & \begin{tabular}[c]{@{}c@{}}-0.1768\\ (0.120)\end{tabular} & \begin{tabular}[c]{@{}c@{}}-0.0234\\ (0.695)\end{tabular} & \begin{tabular}[c]{@{}c@{}}-0.0539\\ (0.699)\end{tabular} & \begin{tabular}[c]{@{}c@{}}0.0128\\ (0.947)\end{tabular} & \begin{tabular}[c]{@{}c@{}}0.0422\\ (0.844)\end{tabular} & \begin{tabular}[c]{@{}c@{}}0.0594\\ (0.777)\end{tabular} \\ \hline
Current\_Effectiveness & \begin{tabular}[c]{@{}c@{}}0.0966\\ (0.189)\end{tabular} & \begin{tabular}[c]{@{}c@{}}0.1559\\ (0.189)\end{tabular} & \begin{tabular}[c]{@{}c@{}}0.0322\\ (0.604)\end{tabular} & \begin{tabular}[c]{@{}c@{}}0.0748\\ (0.607)\end{tabular} & \begin{tabular}[c]{@{}c@{}}0.0540\\ (0.788)\end{tabular} & \begin{tabular}[c]{@{}c@{}}-0.1436\\ (0.518)\end{tabular} & \begin{tabular}[c]{@{}c@{}}-0.1510\\ (0.493)\end{tabular} \\ \hline
Current\_Facial\_Recognition & \begin{tabular}[c]{@{}c@{}}0.0158\\ (0.822)\end{tabular} & \begin{tabular}[c]{@{}c@{}}-0.2495\\ (0.028)*\end{tabular} & \begin{tabular}[c]{@{}c@{}}0.0464\\ (0.436)\end{tabular} & \begin{tabular}[c]{@{}c@{}}0.1725\\ (0.215)\end{tabular} & \begin{tabular}[c]{@{}c@{}}-0.0420\\ (0.827)\end{tabular} & \begin{tabular}[c]{@{}c@{}}0.0022\\ (0.992)\end{tabular} & \begin{tabular}[c]{@{}c@{}}0.0327\\ (0.878)\end{tabular} \\ \hline
Biases\_discrimination & \begin{tabular}[c]{@{}c@{}}-0.0268\\ (0.711)\end{tabular} & \begin{tabular}[c]{@{}c@{}}0.1277\\ (0.276)\end{tabular} & \begin{tabular}[c]{@{}c@{}}-0.0557\\ (0.372)\end{tabular} & \begin{tabular}[c]{@{}c@{}}0.2856\\ (0.054)\end{tabular} & \begin{tabular}[c]{@{}c@{}}-0.1754\\ (0.373)\end{tabular} & \begin{tabular}[c]{@{}c@{}}0.0308\\ (0.889)\end{tabular} & \begin{tabular}[c]{@{}c@{}}-0.0438\\ (0.841)\end{tabular} \\ \hline
Data\_Usage & \begin{tabular}[c]{@{}c@{}}0.1111\\ (0.136)\end{tabular} & \begin{tabular}[c]{@{}c@{}}0.0533\\ (0.656)\end{tabular} & \begin{tabular}[c]{@{}c@{}}0.0585\\ (0.354)\end{tabular} & \begin{tabular}[c]{@{}c@{}}0.0651\\ (0.660)\end{tabular} & \begin{tabular}[c]{@{}c@{}}-0.1111\\ (0.583)\end{tabular} & \begin{tabular}[c]{@{}c@{}}-0.0084\\ (0.970)\end{tabular} & \begin{tabular}[c]{@{}c@{}}-0.0857\\ (0.703)\end{tabular} \\ \hline
System\_Understanding & \begin{tabular}[c]{@{}c@{}}0.0448\\ (0.531)\end{tabular} & \begin{tabular}[c]{@{}c@{}}-0.0145\\ (0.900)\end{tabular} & \begin{tabular}[c]{@{}c@{}}-0.0013\\ (0.983)\end{tabular} & \begin{tabular}[c]{@{}c@{}}0.4410\\ (0.002)*\end{tabular} & \begin{tabular}[c]{@{}c@{}}-0.1470\\ (0.450)\end{tabular} & \begin{tabular}[c]{@{}c@{}}-0.0416\\ (0.848)\end{tabular} & \begin{tabular}[c]{@{}c@{}}0.1141\\ (0.601)\end{tabular} \\ \hline
Privacy\_Threat & \begin{tabular}[c]{@{}c@{}}0.0335\\ (0.639)\end{tabular} & \begin{tabular}[c]{@{}c@{}}-0.3830\\ (0.001)*\end{tabular} & \begin{tabular}[c]{@{}c@{}}0.0173\\ (0.776)\end{tabular} & \begin{tabular}[c]{@{}c@{}}-0.0659\\ (0.642)\end{tabular} & \begin{tabular}[c]{@{}c@{}}0.0229\\ (0.907)\end{tabular} & \begin{tabular}[c]{@{}c@{}}0.0555\\ (0.799)\end{tabular} & \begin{tabular}[c]{@{}c@{}}0.2745\\ (0.201)\end{tabular} \\ \hline
Current\_Law\_Use & \begin{tabular}[c]{@{}c@{}}0.1015\\ (0.225)\end{tabular} & \begin{tabular}[c]{@{}c@{}}0.0981\\ (0.458)\end{tabular} & \begin{tabular}[c]{@{}c@{}}0.0183\\ (0.792)\end{tabular} & \begin{tabular}[c]{@{}c@{}}-0.2543\\ (0.115)\end{tabular} & \begin{tabular}[c]{@{}c@{}}-0.0580\\ (0.798)\end{tabular} & \begin{tabular}[c]{@{}c@{}}0.0587\\ (0.824)\end{tabular} & \begin{tabular}[c]{@{}c@{}}-0.0792\\ (0.742)\end{tabular} \\ \hline
Feel\_Safer & \begin{tabular}[c]{@{}c@{}}0.0491\\ (0.522)\end{tabular} & \begin{tabular}[c]{@{}c@{}}0.3147\\ (0.010)*\end{tabular} & \begin{tabular}[c]{@{}c@{}}0.0157\\ (0.807)\end{tabular} & \begin{tabular}[c]{@{}c@{}}-0.3227\\ (0.030)*\end{tabular} & \begin{tabular}[c]{@{}c@{}}0.0858\\ (0.688)\end{tabular} & \begin{tabular}[c]{@{}c@{}}-0.1444\\ (0.531)\end{tabular} & \begin{tabular}[c]{@{}c@{}}-0.0331\\ (0.885)\end{tabular} \\ \hline
Tech\_Beneficial & \begin{tabular}[c]{@{}c@{}}0.1722\\ (0.026)*\end{tabular} & \begin{tabular}[c]{@{}c@{}}0.3860\\ (0.002)*\end{tabular} & \begin{tabular}[c]{@{}c@{}}0.0621\\ (0.338)\end{tabular} & \begin{tabular}[c]{@{}c@{}}-0.1496\\ (0.325)\end{tabular} & \begin{tabular}[c]{@{}c@{}}0.1334\\ (0.543)\end{tabular} & \begin{tabular}[c]{@{}c@{}}-0.2347\\ (0.307)\end{tabular} & \begin{tabular}[c]{@{}c@{}}-0.0138\\ (0.953)\end{tabular} \\ \hline
Using\_AI & \begin{tabular}[c]{@{}c@{}}0.0626\\ (0.437)\end{tabular} & \begin{tabular}[c]{@{}c@{}}0.0797\\ (0.541)\end{tabular} & \begin{tabular}[c]{@{}c@{}}0.0071\\ (0.916)\end{tabular} & \begin{tabular}[c]{@{}c@{}}-0.0689\\ (0.666)\end{tabular} & \begin{tabular}[c]{@{}c@{}}0.6261\\ (0.005)*\end{tabular} & \begin{tabular}[c]{@{}c@{}}-0.2596\\ (0.299)\end{tabular} & \begin{tabular}[c]{@{}c@{}}0.0575\\ (0.809)\end{tabular} \\ \hline
Tech\_Facial\_Recognition & \begin{tabular}[c]{@{}c@{}}0.0300\\ (0.733)\end{tabular} & \begin{tabular}[c]{@{}c@{}}0.0544\\ (0.700)\end{tabular} & \begin{tabular}[c]{@{}c@{}}-0.0654\\ (0.374)\end{tabular} & \begin{tabular}[c]{@{}c@{}}-0.0691\\ (0.691)\end{tabular} & \begin{tabular}[c]{@{}c@{}}0.0392\\ (0.869)\end{tabular} & \begin{tabular}[c]{@{}c@{}}0.0300\\ (0.909)\end{tabular} & \begin{tabular}[c]{@{}c@{}}-0.0151\\ (0.953)\end{tabular} \\ \hline
Tech\_Law\_Use & \begin{tabular}[c]{@{}c@{}}0.1869\\ (0.022)*\end{tabular} & \begin{tabular}[c]{@{}c@{}}0.0942\\ (0.470)\end{tabular} & \begin{tabular}[c]{@{}c@{}}-0.0259\\ (0.706)\end{tabular} & \begin{tabular}[c]{@{}c@{}}-0.2341\\ (0.139)\end{tabular} & \begin{tabular}[c]{@{}c@{}}0.4693\\ (0.044)*\end{tabular} & \begin{tabular}[c]{@{}c@{}}0.4752\\ (0.066)\end{tabular} & \begin{tabular}[c]{@{}c@{}}0.1124\\ (0.646)\end{tabular} \\ \hline
Notification & \begin{tabular}[c]{@{}c@{}}0.0608\\ (0.464)\end{tabular} & \begin{tabular}[c]{@{}c@{}}0.2847\\ (0.033)*\end{tabular} & \begin{tabular}[c]{@{}c@{}}0.0485\\ (0.486)\end{tabular} & \begin{tabular}[c]{@{}c@{}}-0.1436\\ (0.379)\end{tabular} & \begin{tabular}[c]{@{}c@{}}0.3087\\ (0.209)\end{tabular} & \begin{tabular}[c]{@{}c@{}}0.0348\\ (0.891)\end{tabular} & \begin{tabular}[c]{@{}c@{}}-0.1999\\ (0.417)\end{tabular} \\ \hline
Privacy\_Concern & \begin{tabular}[c]{@{}c@{}}0.0882\\ (0.186)\end{tabular} & \begin{tabular}[c]{@{}c@{}}-0.1170\\ (0.279)\end{tabular} & \begin{tabular}[c]{@{}c@{}}-0.0572\\ (0.310)\end{tabular} & \begin{tabular}[c]{@{}c@{}}0.2415\\ (0.069)\end{tabular} & \begin{tabular}[c]{@{}c@{}}-0.1614\\ (0.385)\end{tabular} & \begin{tabular}[c]{@{}c@{}}-0.0713\\ (0.728)\end{tabular} & \begin{tabular}[c]{@{}c@{}}0.0462\\ (0.818)\end{tabular} \\ \hline
\bottomrule
\end{tabular}
\end{table*}

\section{Discussion} \label{Discussion}
The study aimed to understand how different demographic groups perceive concerns related to public safety, the effectiveness of video surveillance systems, and perceptions of AI-driven video surveillance technology. The research questions formed the foundation for our analysis, and the results provided significant and surprising insights.

\subsection*{RQ1: Do public safety concerns differ across different demographic groups?}

Our findings reveal that public safety concerns indeed vary across different demographic groups. For instance, females, particularly younger ones, are more likely to believe that the presence of law enforcement is insufficient and that crime rates have increased. This aligns with the insight that females feel safer with the current video surveillance (VS) systems in place and have fewer privacy concerns. Moreover, educational level plays a pivotal role in shaping public safety concerns. Those with more education are less concerned about human trafficking, hinting at the role of income and information accessibility in their perceptions. Different ethnic groups also present varied concerns. Latinos generally feel safer than American whites, while African Americans are more concerned about violent crimes compared to the target group. It's essential to understand these differences to target interventions effectively.

\subsection*{RQ2: Do people with different demographics perceive the effectiveness of existing video surveillance systems differently?}

Demographics seem to profoundly influence the perception of the effectiveness of current VS systems. Females, especially the younger demographic, view the current VS as more effective, leading to a heightened sense of security and fewer privacy concerns. This trust, however, comes with a caveat as they are more tolerant of system errors. Older individuals also appreciate the benefits of VS in deterring activities like drug offenses, violent crimes, and vandalism. However, they show less concern towards offenses such as sexual harassment. Among racial groups, Asian individuals, especially the younger cohort, believe that law enforcement effectively addresses public safety, which indicates their trust in existing systems. These patterns suggest that while current VS systems are effective to an extent, there's room for improvement, especially in addressing specific demographic concerns.

\subsection*{RQ3: Does demographic background affect people's perception of AI-driven video surveillance technology?}

Regarding AI-driven video surveillance technology, demographic background undoubtedly plays a significant role. Although older Black individuals are concerned about data usage, they find value in the technology and believe it should be leveraged more extensively. This sentiment contrasts with the general perception of Black respondents who desire a deeper understanding of surveillance systems. Moreover, those with more education, particularly females, are less inclined to believe that law enforcement should use the technology. This could be attributed to a mistrust stemming from concerns about personal information protection. Age plays a distinctive role, with older individuals being more concerned about drug-related crimes and vandalism. On the other hand, females, especially those with a Bachelor's degree, show increased concern about biases and discrimination in technology.

Several other nuanced insights emerged from the study. Older individuals, especially females, are more concerned about venues such as malls but less about streets. This could be attributed to the perceived vulnerability in enclosed spaces. Furthermore, educational level intricately influences the perception of various offenses. For instance, older individuals with higher education are less concerned about streets and more about mall incidents. Their concerns are majorly centered around violent crimes, contrasting the younger generation's heightened sensitivity towards sexual harassment.

Based on these findings, we are proposing the following policy implications: 

\begin{itemize}
    \item \textbf{Gender-based perceptions}: Female respondents, irrespective of other demographic attributes, tend to trust technology more, valuing its safety benefits while showing less concern over privacy. As such, any policy or technological development should emphasize security features and mitigating privacy concerns to appeal more broadly to male respondents.
    
    \item \textbf{Age and Crime Perception}: Younger individuals have varied concerns based on the nature of crimes, while older individuals lean more towards concerns about violent crimes, vandalism, and drug-related issues. This suggests a need for targeted policy measures for different age groups, focusing on their respective concerns. 
    
    \item \textbf{Education and Tech Adoption}: Higher education correlates with a nuanced view of technology's role in law enforcement. While seeing the benefits of the technology, the more educated demographic expresses trust issues, particularly related to personal data handling by law enforcement. Policymakers should consider transparency and public awareness campaigns on how data is used and protected, particularly targeting this demographic.
    
    \item \textbf{Ethnicity and Tech Trust}: Different ethnic groups demonstrate unique perceptions about safety, with specific areas of concern. For example, Blacks and Latinos have varying levels of trust in facial recognition technology (FRT) and its error rates. Policies should be crafted with a culturally competent lens, addressing the unique concerns and sensibilities of each ethnic group.
    
    \item \textbf{Concerns around Biases and Discrimination}: Across many demographics, concerns about biases and discrimination in technology are evident. Policy measures should emphasize the development of unbiased algorithms, regular audits for biases, and public disclosures of these audits to instill greater trust.
    
    \item \textbf{Venue-based Safety Concerns}: Both gender and ethnic groups express concerns about specific venues like malls, streets, and parking lots. Based on these insights, policymakers and law enforcement should prioritize safety measures and patrols, adjusting according to the perceived concerns of different demographic groups.
    
    \item \textbf{Privacy Concerns and Error Tolerance}: While certain demographics like older females express discomfort with higher False Positives, indicating a distaste for excessive notifications, others show a higher tolerance. Policies should strike a balance, optimizing the technology to reduce false alarms while ensuring security. 

\end{itemize}

\section*{Acknowledgment}
This research is supported by the National Science Foundation (NSF) under Award No. 1831795.

\bibliographystyle{cas-model2-names}

\bibliography{cas-refs}

\end{document}